\documentclass[final,5p,times,twocolumn,authoryear]{elsarticle}

\journal{High Energy Astrophysics}

\newcommand{\DMconst}{\,\text{MHz}^2\,\text{pc}^{-1}\,\text{cm}^3\,\text{s}}
\newcommand{\dmu}{\text{pc}\,/\text{cm}^3}

\newcommand{\update}[1]{\textcolor{black}{#1}}
\newcommand{\newupdate}[1]{\textcolor{black}{#1}}

\usepackage{amssymb}
\usepackage{amsmath}
\usepackage{orcidlink}
\usepackage[switch]{lineno} 
\usepackage{graphicx}
\usepackage{subfigure}
\usepackage{bm}
\usepackage{orcidlink}
\usepackage{aas_macros}
\usepackage{subcaption}
\usepackage{subfigure}
\usepackage{cprotect}
\usepackage[svgnames]{xcolor}
\usepackage[inline]{enumitem}
\usepackage{booktabs}
\usepackage{multirow}
\usepackage{nicematrix}
\usepackage{listings}
\usepackage{tabularx}

\definecolor{codegreen}{rgb}{0,0.6,0}
\definecolor{codegray}{rgb}{0.5,0.5,0.5}
\definecolor{codepurple}{rgb}{0.58,0,0.82}
\definecolor{backcolour}{rgb}{0.95,0.95,0.92}
\lstdefinestyle{pystyle}{
  backgroundcolor=\color{backcolour}, commentstyle=\color{codegreen},
  keywordstyle=\color{magenta},
  numberstyle=\tiny\color{codegray},
  stringstyle=\color{codepurple},
  basicstyle=\ttfamily\footnotesize,
  breakatwhitespace=false,         
  breaklines=true,                 
  captionpos=b,                    
  keepspaces=true,                 
  numbers=left,                    
  numbersep=5pt,                  
  showspaces=false,                
  showstringspaces=false,
  showtabs=false,                  
  tabsize=2
}
\newcolumntype{Y}{>{\centering\arraybackslash}X}
\usepackage{afterpage}

\usepackage{hyperref}
\hypersetup{
    colorlinks=true,
}

\begin{document}

\begin{frontmatter}

\title{\update{\textsf{PSRDISP}: A novel approach to modeling dispersive processes in single-pulsar noise analysis using epoch-wise dispersion measures}}

\author[1]{Churchil Dwivedi \orcidlink {0000-0002-8804-650X}}
\affiliation[1]{organization={Astronomy and Astrophysics Division, Physical Research Laboratory},
    addressline={Thaltej Campus, Thaltej}, 
    city={Ahmedabad},
    postcode={380059}, 
    state={Gujarat},
    country={India}}
    
\author[2]{Abhimanyu Susobhanan \orcidlink{0000-0002-2820-0931}}
\affiliation[2]{organization={School of Physics, Indian Institute of Science Education and Research Thiruvananthapuram},
    addressline={Maruthamala PO}, 
    city={Thiruvananthapuram},
    postcode={695551}, 
    state={Kerala},
    country={India}}

\begin{abstract}

\update{We present \textsf{PSRDISP}, a novel approach to modeling deterministic and stochastic dispersive processes in pulsar timing datasets using high-precision epoch-wise dispersion measure (DM) estimates, with a Gaussian Process-based approach. Unlike the conventional single-pulsar noise analysis methodology, which is applied to frequency-resolved times of arrival (ToAs) of pulses, this technique is applied to epoch-wise DMs which are derived from these ToAs. It can also be applied to wideband DMs measured simultaneously with wideband ToAs. Therefore, this framework provides a paradigm-agnostic approach to characterise single-pulsar dispersive processes. This method is expected to minimise the impact of achromatic red noise processes while characterising these dispersive effects. We substantiate the discussed technique with representative examples using simulated narrowband and wideband datasets with realistic noise injections. We found the recovery to be in close agreement with the injections, and agnostic to the estimation technique.} Our method applies to pulsar timing experiments where precise, epoch-wise DM estimates are possible, such as the Indian Pulsar Timing Array. This technique can serve as a powerful diagnostic tool for validating single-pulsar noise analyses, which is crucial for precision pulsar timing experiments, such as Pulsar Timing Arrays.

\end{abstract}


\end{frontmatter}

\section{Introduction}\label{sec:1}

Millisecond Pulsars (MSPs) are rotating neutron stars with millisecond-scale periods, acting as accurate celestial clocks due to their high rotational stability \citep{LorimerKramer2004, HobbsGuo+2019}. Pulsar timing is the technique of tracking a pulsar's rotation by measuring the times of arrival (ToAs) of its pulses \update{\citep{HobbsEdwardsManchester2006, Edwards+2006}}. It is one of the most precise techniques in astrophysics, enabling us to study a wide range of time-domain phenomena ranging from testing strong-field gravity \citep{KramerStairs+2021} to coronal mass ejections \citep{ChowdhuryKrishnakumar+2026}. Recently, Pulsar Timing Array \citep[PTA:][]{FosterBacker1990} \newupdate{experiments reported evidence for a stochastic Gravitational Wave Background (GWB)} in the nanohertz frequency range using precise timing of an ensemble of MSPs \citep{Agazie+2023, Reardon+2023, Antoniadis+2023c, Xu+2023, AgazieAntoniadis+2024, Miles+2025b}.

The propagation of electromagnetic waves from a pulsar through the ionized interstellar medium (IISM) and the interplanetary medium introduces chromatic delays, primarily due to the frequency-dependent refractive index of the medium. Assuming a homogeneous IISM containing cold, ionised plasma, in the absence of Galactic magnetic fields and finite temperature effects \citep{Kulkarni2020}, we can write the refractive index as \citep{LorimerKramer2004}
\begin{equation}\label{eq:1.1}
    \mu=\sqrt{1-\left(\frac{\nu_p}{\nu}\right)^2}
\end{equation}
where $\nu_p$ is the plasma frequency, given by
\begin{equation} \label{eq:plasmafreq}
\nu_p=\sqrt{\frac{e^2n_e}{\pi m_e}}
\end{equation}
Here, $n_e$ is the free electron number density in the medium, and $e$ and $m_e$ are respectively the charge and mass of an electron. Due to this frequency dependence of $\mu$, a radio wave of frequency $\nu$ will experience a propagation delay while traveling a distance $d_L$ relative to an infinite-frequency signal, given by
\begin{equation}\label{eq:1.2}
    \Delta t=\left(\int_0^{d_L}\frac{\mathrm d\ell}{V}\right)-\frac{d_L}{c}
\end{equation}
where $V=c\mu$ is the speed of the radio wave in the IISM and $c$ is the speed of light in vacuum. Using equation \eqref{eq:1.1}, assuming $\nu_p\ll\nu$, and evaluating the integral in equation \eqref{eq:1.2}, we obtain the dispersion delay caused by a homogeneous IISM as
\begin{equation}\label{eq:1.3}
    \Delta t=\frac{\mathcal{D}\times\rm DM}{\nu^2}
\end{equation}
where
\begin{equation}\label{eq:1.4}
    \mathrm{DM} = \int_0^{d_L} n_e \,\mathrm{d}\ell
\end{equation}
is the integrated free electron column density along the line of sight, and is called the Dispersion Measure (DM), and $\mathcal{D}\approx4.1488\times10^3 \DMconst$ is known as the dispersion constant \citep{LorimerKramer2004}. 

\update{In general, the DM of a pulsar is not constant; it varies with time depending on the line of sight due to the relative motion between the pulsar and the Earth, and the inherently turbulent nature of the IISM.} These temporal variations are usually modeled by pulsar timing packages, such as \newupdate{\textsf{TEMPO2} \citep{HobbsEdwardsManchester2006}} and \textsf{PINT} \citep{Luo+2021, Susobhanan+2024}, using a low-order Taylor series expansion in time around a fiducial epoch (referred to as the \texttt{DM\_Taylor} model). Such a model can account for the \newupdate{slow variations} in DMs, but it falls short in modeling the stochasticity arising due to random fluctuations occurring on varying timescales due to the intrinsic variability of the IISM, and the relative motion between the Earth and the pulsar \citep[e.g.][]{DonnerVerbiest+2020}. These stochastic DM variations, known as DM noise (hereafter \texttt{DMN}), are ubiquitous in pulsar timing and different approaches have been adopted over the years to model them in the measured ToAs. These include the \texttt{DMMODEL} approach of \citet{KeithColes+2012}, which provides a spline-based representation of DM variations, the \texttt{DMX} approach described in \citet{ArzoumanianBrazier+2015}, where DM variations are modeled as a piecewise-constant function in time, the \texttt{DMGP} approach of \citet{LentatiAlexander+2014}, which represents the DM variations as a reduced-rank Fourier-domain Gaussian Process (GP), typically with a power-law spectrum, and the \texttt{DMWaveX} approach in PINT \citep{Susobhanan+2024}, which uses an unconstrained Fourier series representation of DM variations. 

Apart from the IISM-induced variations, the solar wind (SW), which is a stream of charged particles originating from the Sun, also alters the electron column density along the line of sight and hence produces variations in the DM. These variations, however, depend on the solar elongation\footnote{The angle between the Sun-Earth and the Earth-pulsar position vectors is called solar elongation. Due to Earth's motion around the Sun, the solar elongation of a pulsar changes with a periodicity of $1\,\rm yr$.} throughout the year and are highest near solar conjunctions \citep{Tiburzi+2019, Tiburzi+2021}. \update{These deterministic variations are generally modeled using a spherically symmetric model of SW \citep{Edwards+2006} with a constant electron density $\bar{n}_e^{\rm sw}$, sometimes with higher-order derivatives to incorporate additional effects \citep{Nobleson+2026}. The SW itself exhibits inherent stochasticity, acting as a source of chromatic noise known as the solar wind noise (hereafter \texttt{SWN}) in pulsar timing. Similar to the \texttt{DMGP} approach, the recent \texttt{SWGP} approach of \citet{Susarla+2024} can be used to account for these effects. Additionally, the \texttt{SWX} model \citep{Agazie+2025} can also be used, where the variations in $n_e^{\rm sw}$ are modeled using a piece-wise constant function in time, similar to the \texttt{DMX} model.}

\update{Another effect caused by the turbulence in the IISM is the variation of $n_e$ on a wide range of length scales}, leading to multi-path propagation of radio waves, that manifests as the broadening of an otherwise sharp intrinsic pulse profile. These distortions produce an apparent delay that scales as 
\begin{equation}\label{eq:1.7}
    \Delta t_{\rm scat}\propto \nu^{-4}
\end{equation}
for Gaussian inhomogeneities\footnote{The actual chromatic index for scattering-induced delays depends strongly on the turbulence spectrum along the line of sight to a pulsar. The $\nu^{-4}$ form is an approximation for a large number of randomly distributed thin scattering screens. For the case of a Kolmogorov turbulence spectrum, these delays scale as $\nu^{-4.4}$.} \citep{LorimerKramer2004}. \update{These profile-shape distortions also experience stochastic variations, giving rise to variable scattering in MSPs \citep{Kulkarni+2025}. These variations are a source of chromatic red noise, commonly modeled in ToAs via GPs \citep{Srivastava+2023} for a given turbulence spectrum. A more generic approach featuring a free-chromatic index noise model ($\Delta t\propto\nu^{-\chi}$) has also been used \citep{Nobleson+2026, Larsen+2026}.} The presence of scatter-broadening introduces a source of bias in the estimated DMs \citep{Singha+2024}, the imprint of which over DM is not straightforward to interpret. \update{These distortions produce a pulsar-dependent non-linear frequency dependence between scattering and other dispersive processes, which poses a roadblock in deducing their inter-dependence}. Profile-domain techniques are, therefore, paramount for estimating scatter broadening \citep{Singha+2024} and for subsequently mitigating its effect from the pulse profiles \citep{Singha+2024, Bathula+2025}. Due to all these inherent challenges, we restrict ourselves to the treatment of purely dispersive processes in the present work.

For high-precision pulsar timing experiments, such as PTAs, these noise processes pose serious challenges. In particular, the presence of long-timescale DM variations can degrade a PTA experiment's ability to accurately characterise gravitational wave signals, especially if they are mis-modeled. \update{Furthermore, mis-modeled solar wind effects can also mimic a Common Red Noise (CRN) process due to a common periodicity for all pulsars in the ensemble \citep{Tiburzi+2016, Susarla+2024}.} Since the starting point of any PTA experiment to detect a GWB is to look for a CRN process in its ensemble of pulsars, it is of paramount importance to accurately characterise the single-pulsar noise processes, so as to prevent any false-positives \citep{Chen+2021, Arzoumanian+2020, Goncharov+2021, Antoniadis+2022}. 

\update{In this work, we address the precise modeling of dispersive noise processes by presenting an alternative methodology, adapting the Fourier-domain GP-based approach of \citet{LentatiAlexander+2014} to model the stochastic effects owing to its favored use in the recent literature \citep{Srivastava+2023, Antoniadis+2023b, Larsen+2024, Miles+2025, Nobleson+2026}. This method directly uses high-precision epoch-wise DM time series to fit for various dispersive noise processes in a purely Bayesian framework. This is in contrast to what is done in conventional single-pulsar noise analysis (SPNA) techniques, which primarily work on ToA residuals and model the delays introduced by various noise processes using packages such as \textsf{ENTERPRISE} \citep{Ellis+2020, Johnson+2024}. Our treatment, therefore, ensures minimal contamination from achromatic red noise and white noise processes, aiding precise estimation of dispersive processes.}

The rest of the paper is organised as follows. In Section \ref{sec:2}, we give a detailed description of the new technique. We provide the details of the simulation framework adopted for the injection and recovery study, along with the results for a representative case to demonstrate the effectiveness of the technique in recovering deterministic and stochastic processes in Section \ref{sec:3}. We discuss the results and their implications in Section \ref{sec:4}, followed by a brief summary in Section \ref{sec:5}, and outline the potential areas of future work in Section \ref{sec:6}.

\section{Description of the technique} \label{sec:2}

In this section, we provide a detailed description of the theoretical construct of this work. It should be noted that the DM measurements considered in this work correspond to high-precision epoch-wise DM estimates. \update{These can be obtained either in the narrowband paradigm using epoch-wise ToA fitting techniques such as \textsf{DMcalc} \citep{Krishnakumar+2021}, or in the wideband paradigm using \textsf{PulsePortraiture} \citep{Pennucci+2014, Pennucci2019}. Additionally, we assume that the DM measurements for different epochs are independent, i.e, the measurements of one epoch do not affect the others\footnote{This makes sure that the measurement covariance matrix, $\bm{\zeta}$, is diagonal.}. This also means that the results of a global \texttt{DMX} or \texttt{DMMODEL} fit should be used with caution with this technique, since such measurements can have non-zero measurement covariance between different epochs. That said, the framework presented in this section is agnostic to the DM estimation algorithms or pulsar timing paradigms (narrowband or wideband) employed, as long as the above mentioned assumptions hold good.}

\newupdate{While DM variations for most pulsars are expected to be induced through interstellar turbulence with a fully stochastic spectrum, in our treatment we follow the common practice in the pulsar timing community and model these variations with a hybrid analytical approach based on both, polynomial terms in the time domain, and a spectral component in the Fourier domain. Likewise, the temporal variations in SW delays affecting pulsar signals are due to fully stochastic turbulent processes, but we model them through combined polynomial and spectral terms. Therefore, we can write the total DM for any pulsar, in the absence of additional scattering-induced effects, as}
\begin{equation}\label{eq:2.1}
\begin{aligned}
    \rm DM(t) \simeq \,\, &\rm DM_0 + \Delta\rm DM^{\rm Pol}(t) + \Delta\rm DM_{\rm SW}^{\rm Pol}(t)  \\
    &+ \Delta\rm DM_{\rm DMN}(t) + \Delta\rm DM_{\rm SWN}(t) + \Delta\rm  DM_{\rm WN}
\end{aligned}
\end{equation}
where $\rm DM_0$ is a constant (offset) DM term, $\Delta\rm DM^{\rm Pol}(t)$ and $\Delta\rm DM_{\rm SW}^{\rm Pol}(t)$ are the contributions from the \newupdate{polynomial} DM and SW models, respectively, $\Delta\rm DM_{\rm DMN}(t)$ and $\Delta\rm DM_{\rm SWN}(t)$ are the contributions from the \texttt{DMN} and \texttt{SWN} stochastic noise processes, respectively, $\Delta\rm DM_{\rm WN}$ is an additional term related to the measurement uncertainty in the DMs, and $t$ represents time. In the present work, we consider a low-order Taylor series expansion in time for $\Delta\rm DM^{\rm Pol}$, which can be written as
\begin{equation}\label{eq:2.2}
    \Delta\rm DM^{\rm Pol} = \rm{DM}_1\left(t-t_{\rm ref}^{\rm DM}\right)+\frac{1}{2!} \rm{DM}_2\left(t-t_{\rm ref}^{\rm DM}\right)^2
\end{equation}
where $\rm{DM}_1$ and $\rm{DM}_2$ respectively represent the first and second order time derivatives of the DM, and $t_{\rm ref}^{\rm DM}$ is the DM reference epoch (referred to as \texttt{DMEPOCH} in pulsar timing). We also consider a similar \newupdate{polynomial} model for the SW process, which can be written, up to second order, as
\begin{equation}\label{eq:2.3}
    \rm SW_{\rm Pol}(t) = \bar{n}_e^{\rm sw}+\left.\frac{\partial n_e^{\rm sw}}{\partial t}\right|_{t=t_{\rm ref}}(t-t_{\rm ref})+\frac{1}{2!}\left.\frac{\partial^2 n_e^{\rm sw}}{\partial t^2}\right|_{t=t_{\rm ref}}(t-t_{\rm ref})^2
\end{equation}
\begin{equation}\label{eq:2.4}
    \Delta\rm DM_{\rm SW}^{\rm Pol}(t) = \rm SW_{\rm Pol}(t)\,\mathcal{G}(\rho, \bm{r}, t)
\end{equation}
where $\rm\bar{n}_e$ represents the constant average solar wind electron number density\cprotect\footnote{Hereafter, the terms (in leading order) in the deterministic SW model will be represented by $\verb|NE_SW|$, $\verb|NE_SW1|$ and $\verb|NE_SW2|$, respectively, to stay coherent with pulsar timing conventions.} at $1\rm AU$, and $t_{\rm ref}^{\rm SW}$ is the solar wind reference epoch (referred as \texttt{SWEPOCH} in pulsar timing). The multiplicative factor $\mathcal{G}$ is a geometric factor which depends on the solar elongation $\rho$ and the position vector of the observatory relative to the Solar System Barycenter (SSB) $\bm r$, and is given by \citep{Edwards+2006, Hazboun+2022}
\begin{equation}\label{eq:2.4b}
    \mathcal{G}(\rho, \bm r, t)=\frac{\rho(t)}{|\bm r|\sin\rho(t)}\rm AU^2
\end{equation}
The quantities $\rho$ and $\bm r$ are computed using solar-system ephemerides, such as \texttt{DE440} \citep{Park+2021}. We can write the stochastic variations induced by the \texttt{DMN} process in terms of a Fourier-basis GP as
\begin{equation}\label{eq:2.5}
    \begin{aligned}
        \Delta\rm DM_{\rm DMN}(t)=\sum_{j=1}^{N_{\rm harm}}&\left(\mathcal{A}_j^{\rm DM}\cos\left[2\pi f_j\left(t-t_{\rm ref}^{\rm DM}\right)\right]\right. \\
        &+\left.\mathcal{B}_j^{\rm DM}\sin\left[2\pi f_j\left(t-t_{\rm ref}^{\rm DM}\right)\right]\,\right)
    \end{aligned}
\end{equation}
where $\{\mathcal{A}^{\rm DM}\}$ and $\{\mathcal{B}^{\rm DM}\}$ are the Fourier noise amplitudes, $f_j=j/T_{\rm span}$ is the Fourier frequency, and $N_{\rm harm}$ is the maximum number of Fourier bins used for the process. It should be noted that, unlike the conventional ToA-based GP implementations such as in \textsf{ENTERPRISE}, these amplitudes have units of DM instead of time. Therefore, in order to ensure that they are comparable with the conventional noise amplitudes, we perform a scaling transformation as
\begin{equation} \label{eq:2.6}
    \mathcal{A}_j^{\rm DM} \to \mathcal{A}_j^{\rm DM}\left(\frac{\nu_{\rm ref}^2}{\mathcal{D}}\right)\,,\,\,\,\mathcal{B}_j^{\rm DM} \to \mathcal{B}_j^{\rm DM}\left(\frac{\nu_{\rm ref}^2}{\mathcal{D}}\right)
\end{equation}
In a similar fashion, we include the stochastic variations induced by the \texttt{SWN} process in terms of a Fourier-basis GP, following the prescription in \citet{Susarla+2024}, as
\begin{equation}\label{eq:2.7}
    \begin{aligned}
        \Delta\rm DM_{\rm SWN}(t)=&\sum_{j=1}^{N_{\rm harm}}\left(\mathcal{A}_j^{\rm SW}\cos\left[2\pi f_j\left(t-t_{\rm ref}^{\rm SW}\right)\right]\right. \\
        &+\left.\mathcal{B}_j^{\rm SW}\sin\left[2\pi f_j\left(t-t_{\rm ref}^{\rm SW}\right)\right]\,\right)\,\mathcal{G}(\rho, \bm r, t)
    \end{aligned}
\end{equation}
where the noise amplitudes $\{\mathcal{A}^{\rm SW}\}$ and $\{\mathcal{B}^{\rm SW}\}$ have the units of $\rm{cm}^{-3}$ as per equation \eqref{eq:2.4}\footnote{The \texttt{SWN} implementation in \citet{Susarla+2024} does not require the further transformation of the \texttt{SWN} Fourier noise amplitudes, unlike as done for \texttt{DMN} using equation \eqref{eq:2.6}. We follow the same prescription to remain coherent with the existing pulsar timing frameworks for better compatibility and cross-validation.}. \update{In order to take care of any systematics associated with the estimated DMs, and compensate for the under/over-estimation of their measurement uncertainties, we use a white-noise process, $\Delta\rm DM_{\rm WN}$, with a standard deviation, $\varsigma_{\rm WN}$, depending on the DM measurement uncertainty $\sigma_{\rm DM}$, given by}
\begin{equation}\label{eq:2.8}
    \varsigma_{\rm WN}^2 = \verb|DISPEFAC|^2\left(\sigma_{\rm DM}^2+\verb|DISPEQUAD|^2\right)
\end{equation}
\cprotect\update{where $\verb|DISPEFAC|$ and $\verb|DISPEQUAD|$ are white-noise parameters employed to account for such systematics. It should be noted that in the case of narrowband timing,  $\varsigma_{\rm WN}$ can be thought of as originating from the error propagated from the ToA measurement uncertainties. Hence, any inaccuracy in the ToA uncertainties (which are accounted by white noise parameters such as \texttt{EFAC}s and \texttt{EQUAD}s) have to be propagated to $\varsigma_{\rm WN}$ also. Throughout this work, we will only consider $\verb|DISPEFAC|$ and set $\verb|DISPEQUAD|=0.0$ for simplicity.}

We can now write the Bayes' Theorem in this context as
\begin{equation}\label{eq:2.9}
    \mathcal{P}(\bm\theta\,|\,\bm d,M)=\frac{\mathcal{L}(\bm d\,|\,\bm\theta,M)\,\Pi(\bm\theta\,|\,M)}{\mathcal{Z}(\bm d\,|\,M)}
\end{equation}
where $\bm\theta$ is the parameter vector, $\bm d$ is the data (containing the DM measurements, their uncertainties, and the corresponding epochs), and $M$ is the model hypothesis. The quantity $\mathcal{P}(\bm\theta\,|\,\bm d,M)$ is the posterior distribution of parameters $\bm\theta$ given the data and the model hypothesis, $\mathcal{L}(\bm d\,|\,\bm\theta,M)$ is the likelihood of the data, given the underlying model hypothesis, $\Pi(\bm\theta\,|\,M)$ is the prior distribution of the model parameters, and $\mathcal{Z}(\bm d\,|\,M)$ is the Bayesian evidence calculated over the prior volume. We can define the log-likelihood function for the DMs in the form of a multivariate Gaussian distribution, similar to the conventional SPNA \citep{LentatiAlexander+2014, Srivastava+2023, Nobleson+2026} approach, as
\begin{equation} \label{eq:2.10}
    \ln\mathcal{L}=-\frac{1}{2}\bm{\eta}^T\bm{\zeta}^{-1}\bm{\eta}-\frac{1}{2}\ln[\det(2\pi\bm{\zeta})]
\end{equation}
where $\bm{\eta}$ is a vector of the DM time series, encapsulating all the deterministic and stochastic processes mentioned in equation \eqref{eq:2.1}, and $\bm{\zeta}$ is the diagonal DM white noise covariance matrix containing the scaled DM uncertainties given by equation \eqref{eq:2.8}. By imposing Gaussian priors on the GP noise amplitudes with a covariance matrix $\bm \Phi(\bm{\theta}_s)$ and hyper-parameters $\bm{\theta}_s$, we can analytically marginalise the likelihood expression following the prescription in \citet{vanHaasteren+Levin2013} to arrive at the marginalised likelihood, given by
\begin{equation}\label{eq:2.11}
    \ln\Lambda=-\frac{1}{2}\left[\bm{\eta}-\Delta_{\rm det}(\bm{\theta}_d)\right]^T\bm{\Xi}^{-1}\left[\bm{\eta}-\Delta_{\rm det}(\bm{\theta}_d)\right]-\frac{1}{2}\ln[\det(2\pi\bm{\Xi})]
\end{equation}
where $\Delta_{\rm det}$ encapsulates all the deterministic processes and $\bm{\theta}_d$ is a vector containing all the respective model parameters. The effective noise covariance matrix, $\bm \Xi$, can be represented in the reduced-rank approximation \citep{LentatiAlexander+2014, vanHaasteren+Vallisneri2014} as
\begin{equation}\label{eq:2.12}
    \bm{\Xi}=\bm{\zeta}+\bm{\varphi\,\Phi\,\varphi}^T
\end{equation}
Here, $\bm \varphi$ is the noise basis matrix that contains, in general, the partial derivatives of DM time series with respect to the deterministic and stochastic components, and can be represented as
\begin{equation}\label{eq:2.13}
    \varphi_{ij}=\frac{\partial \eta_i}{\partial a_j}\;,\;\forall\; a\in \left\{\bm{\theta}_d\right\}\cup\left\{\bm{\mathcal{A}^{\rm DM}}, \bm{\mathcal{B}^{\rm DM}}\right\}\cup\left\{\bm{\mathcal{A}^{\rm SW}}, \bm{\mathcal{B}^{\rm SW}}\right\}
\end{equation}
where $\left\{\bm{\mathcal{A}^{\rm DM}}, \bm{\mathcal{B}^{\rm DM}}\right\}$ and $\left\{\bm{\mathcal{A}^{\rm SW}}, \bm{\mathcal{B}^{\rm SW}}\right\}$ represent the set of Fourier components for the \texttt{DMN} and \texttt{SWN} processes, respectively. The covariance matrix of Fourier amplitudes, $\bm{\Phi}$, is related to the Fourier-domain power-spectral density (PSD), $S(f)$, of the noise processes as
\begin{equation} \label{eq:2.14}
    \bm{\Phi}_{n\,m\,\alpha\,\beta}=S\left(f_n;\left\{\bm{\theta}_s\right\}_{\alpha}\right)\,\delta_{nm}\,\delta_{\alpha\beta}\,/\,T_{\rm span}
\end{equation}
where $n,\,m=1,2,3,\dots,N_{\rm harm}$, represent the Fourier basis indices, $\alpha,\,\beta=1,2,3,\dots,N_{\rm psr}$, represent the pulsar indices for a PTA, $\delta$ is the Kronecker-delta function, and the PSD is given by
\begin{equation} \label{eq:2.15}
    S(f)=\frac{A^2}{12\pi^2}\left(\frac{f}{f_{\rm yr}}\right)^{-\gamma}\left(\frac{\nu}{\nu_{\rm ref}}\right)^{-2\chi}\rm yr^3
\end{equation}
where $f_{\rm yr}$ is the $1\,\rm{yr}^{-1}$ reference Fourier frequency, $\nu_{\rm ref}$ is taken as $1400\,\rm MHz$ and $\chi=2$ for dispersive processes. In a more advanced treatment, we can consider solar wind as a spatially-correlated process, but this is beyond the scope of the current work. It is worthwhile to mention that in the implemented technique, we also analytically marginalise over the DM offset and $\Delta\rm DM^{\rm Pol}$ parameters, such that $\bm{a}$ not only contains the Fourier noise amplitudes, but also includes the $\rm{DM}_0$, $\rm{DM}_1$ and $\rm{DM}_2$ parameters. The corresponding diagonal entries in $\bm{\Phi}$ will be infinite. This methodology is also consistent with the implementation in \textsf{ENTERPRISE}. Thus, equation \eqref{eq:2.9} can be re-written in terms of the marginalised likelihood, $\Lambda$, to sample the reduced parameter space comprising of $\bm{\theta}_s$ and $\bm{\theta}_d$. We use the \textsf{emcee} package \citep{Foreman-Mackey+2013} to sample the posterior distributions of parameters with an affine-invariant Markov Chain Monte-Carlo (MCMC) sampling technique \citep{Metropolis+1953, Hastings1970, Earl+2005, GoodmanWeare2010} using the priors mentioned in Table \ref{tab:prior-distributions}. To assess the convergence, we visually inspect the MCMC chains after discarding about 25\% of the initial samples for all the fitted parameters.

\update{In the rest of the paper, we will denote the methodology described in this section as the \textsf{PSRDISP} technique. At this juncture, it is worthwhile to point out the uniqueness of this technique compared to the \textsf{ENTERPRISE} framework, or even the work presented in \citet{Susobhanan+vanHaasteren2025} for the wideband paradigm. The \textsf{ENTERPRISE} implementation models noises in terms of delays introduced in the ToAs. This means that both chromatic and achromatic processes are kept on an equal footing. Such an implementation is susceptible to potential spectral leakages, primarily due to modeling all red noises in the Fourier-domain via GPs \citep{Iraci+2024}. The treatment in \citet{Susobhanan+vanHaasteren2025} uses simultaneous ToAs and DMs estimated via a global fit in the wideband paradigm, however, still modeling the processes partly in terms of delays. \textsf{PSRDISP}, on the other hand, directly models the aforementioned dispersive processes in terms of the $\delta\rm{DM}$ produced \newupdate{ by directly constructing the noise covariance matrices through these $\delta\rm{DM}$s. This makes this approach unique in the sense that no ToA delays are incorporated at any point to correct for these effects}. Such a treatment is almost unaffected by achromatic red noise, which gives the potential to accurately characterise the chromatic noises. Therefore, this framework provides a \newupdate{complementary} domain for modeling these processes and give \textit{an independent check} on the conventional methods.}

\section{A representative example} \label{sec:3}

\begin{figure*}[!ht]
    \centering
    \includegraphics[trim=0cm 0cm 1.5cm 0cm, clip, width=\linewidth]{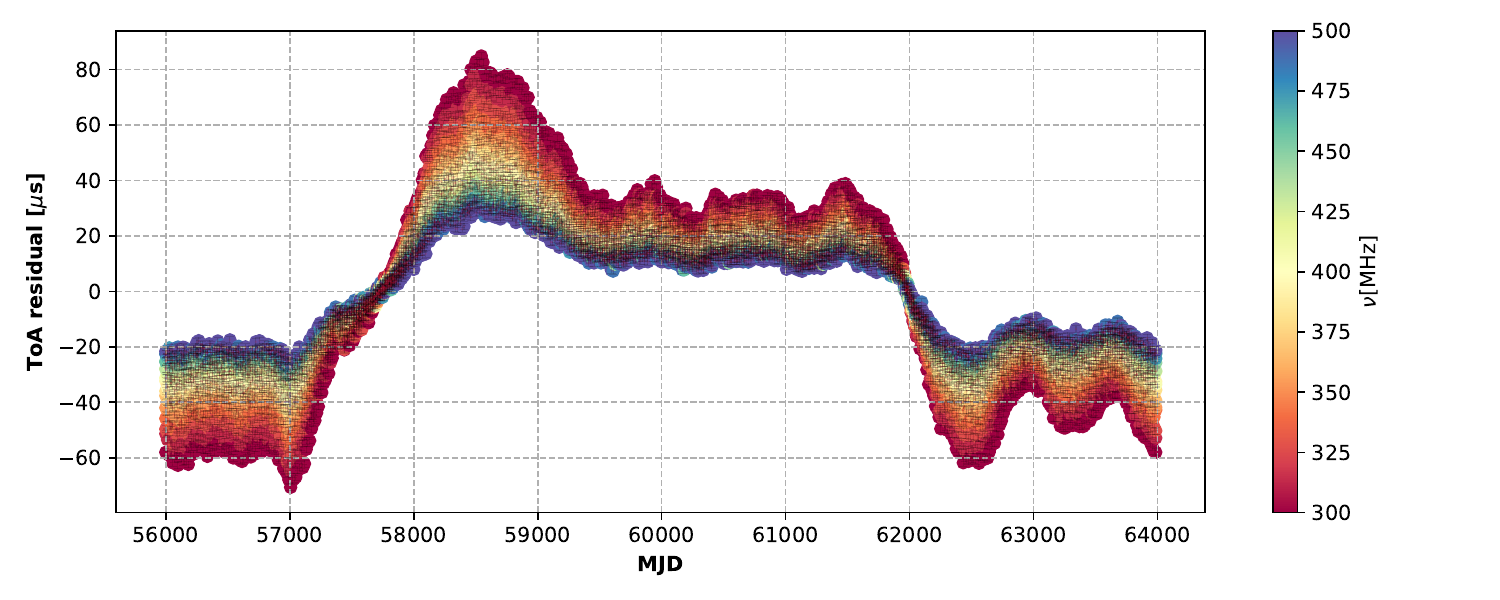}
    \cprotect\caption{The simulated narrowband ToA residuals with 16 sub-bands in the $300-500\,\rm{MHz}$ observing frequency range. The ToAs have a sensitivity of $1\mu\rm s$. The deterministic SW process is simulated with $\verb|NE_SW|=2.0$, $\verb|NE_SW1|=1.0$, and $\verb|NE_SW2|=0.75$, while the DM process is set as per the original timing solution. The \texttt{DMN} process is injected with $\log_{10}A=-13.5$ and $\gamma=3.0$. Additional 16 Fourier bins below the $1/T_{\rm span}$ bin are incorporated to simulate a realistic noise process. The high-frequency cutoff is set to $1000/T_{\rm span}$.}
    \label{fig:ref-DM-det-dmn-arn-swn-injections-narrowband}
\end{figure*}

\begin{figure*}[!ht]
\centering
    \subfigure[DM time series with recovered noise realisations]{\includegraphics[trim=0 0 0 0, clip, width=0.49\textwidth]{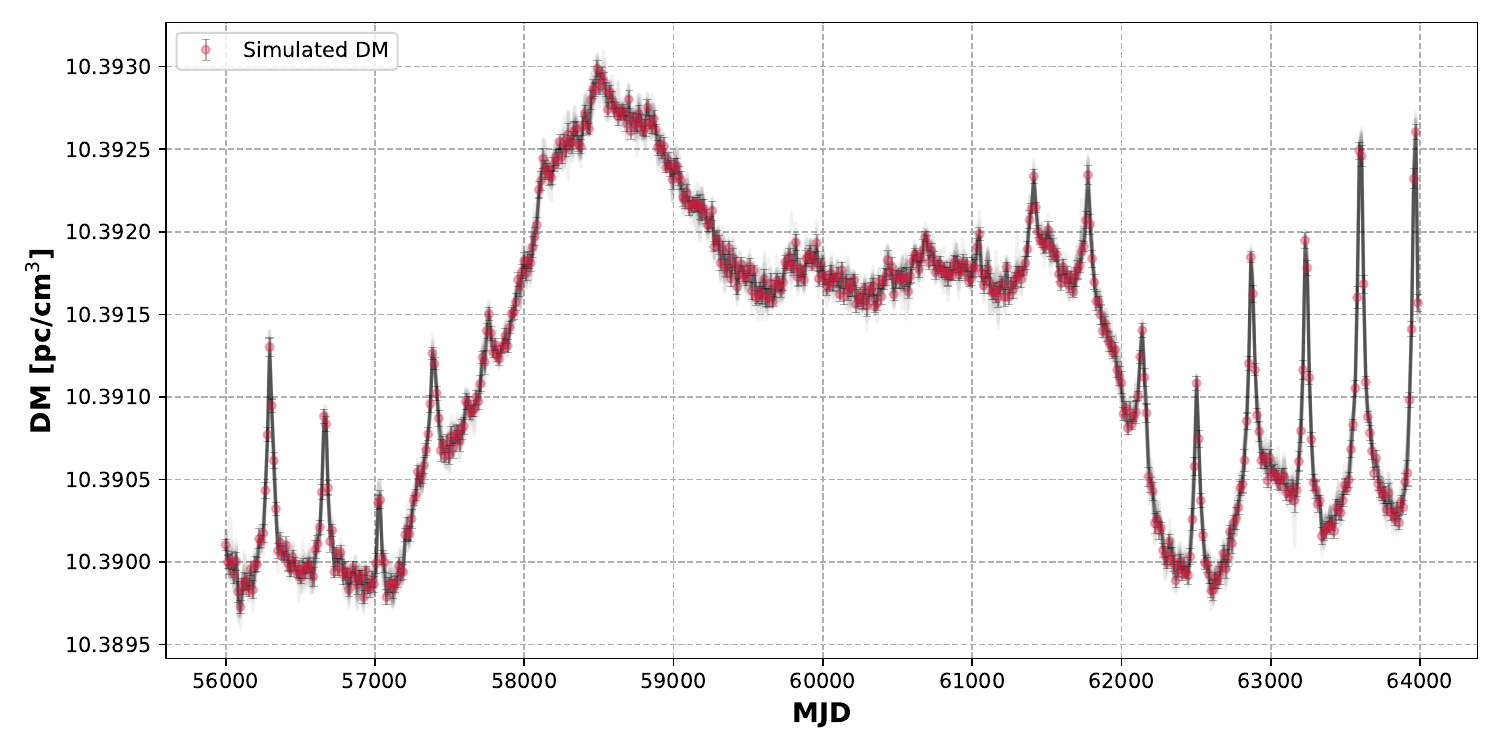}\label{fig:rev-DM-det-dmn-arn-swn-realisations-narrowband}}
    \subfigure[Injected and recovered noise realisations]{\includegraphics[trim=0 0 0 0, clip, width=0.49\textwidth]{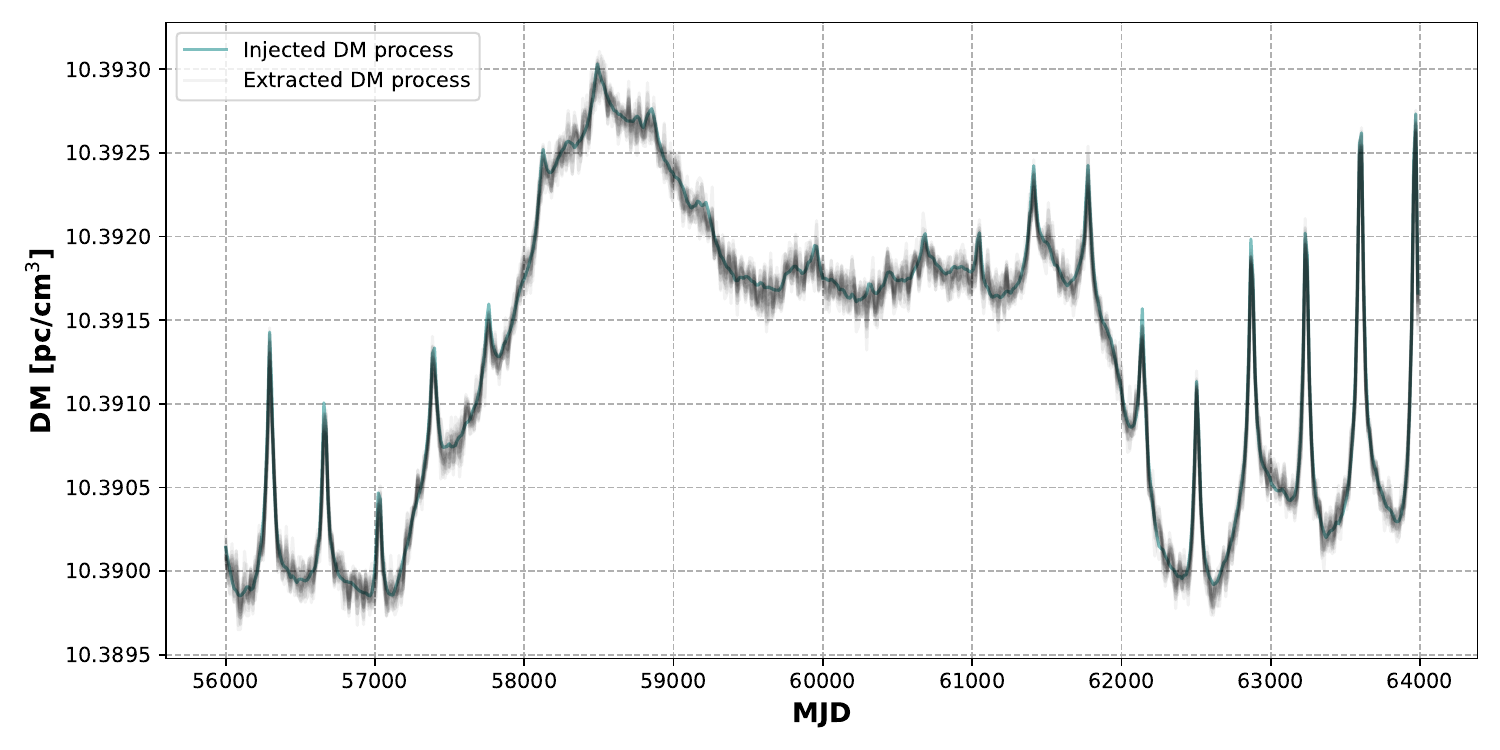}\label{fig:rev-DM-det-dmn-arn-swn-realisations-injection-narrowband}}
\cprotect\caption{The DM time series extracted from the simulated narrowband ToAs (as crimson points) with 16 sub-bands using the \textsf{DMcalc} package, along with 20 random time-domain reconstructed noise process realisations using parameter estimates obtained after applying the \textsf{PSRDISP} technique on the DM time series (shown in solid black), shown in the left panel. The median precision of the estimated DMs is $\sim4.6\times10^{-5}\,\dmu$. The right panel shows the injected DM process realisation (teal) compared with the 20 random recovered DM process realisations (black).}
\end{figure*}

\begin{figure*}[!ht]
    \centering
    \includegraphics[trim=0 0 0 0, clip, width=\textwidth]{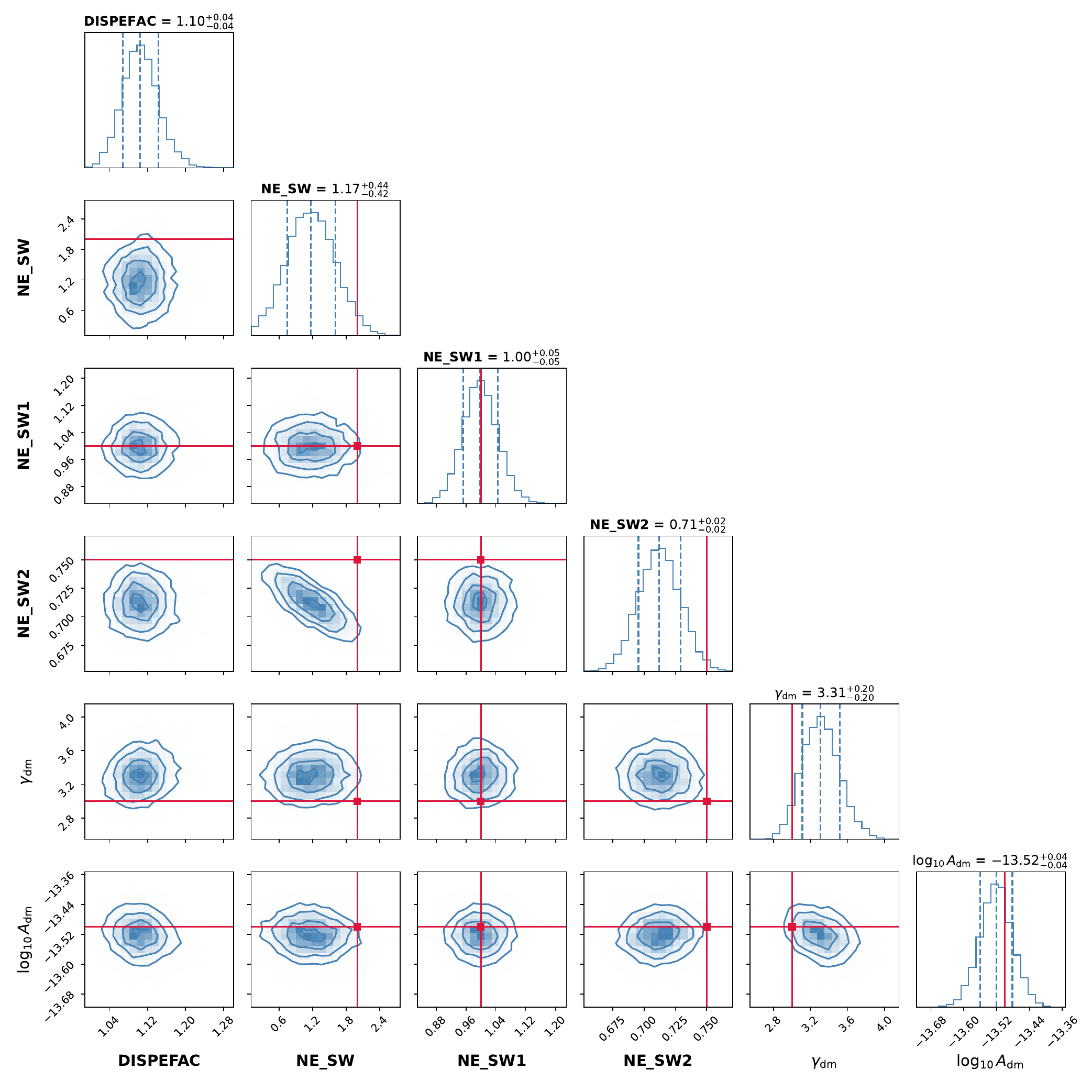}
    \cprotect\caption{The posterior distributions of the deterministic SW process and the DM \texttt{WN} parameters, along with \texttt{DMN} hyper-parameters, estimated using the \textsf{PSRDISP} technique on the DM time series extracted from the simulated narrowband ToAs. The injected values of various parameters are highlighted as crimson vertical lines, while the median values along with $1\sigma$ ranges are shown at the top of each posterior panel.}
    \label{fig:rev-DM-det-dmn-arn-swn-noise-narrowband}
\end{figure*}

\begin{figure*}[!ht]
    \centering
    \includegraphics[trim=0 0 0 0, clip, width=\textwidth]{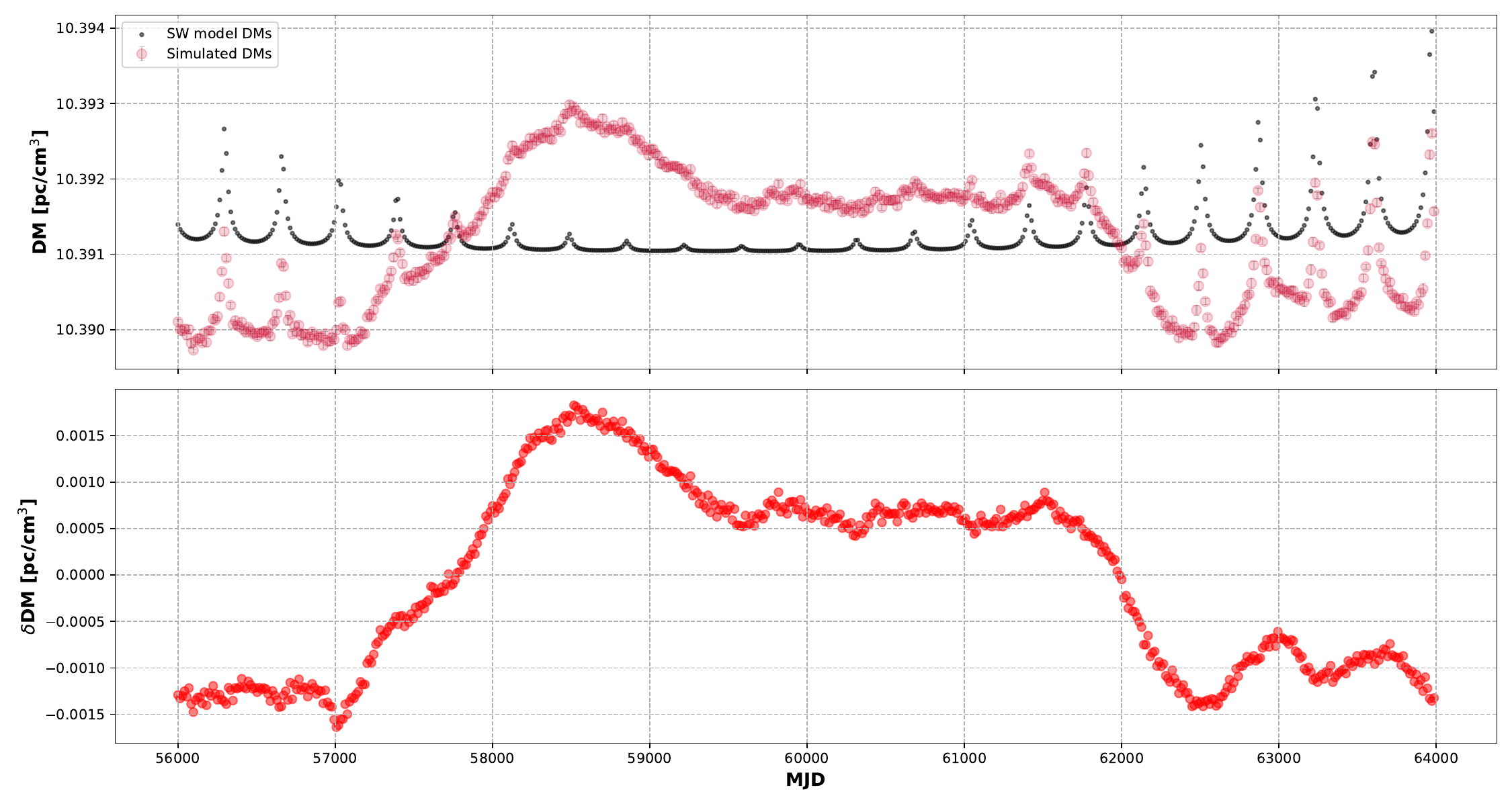}
    \cprotect\caption{The extracted DM time series from narrowband ToAs along with the deterministic SW process estimated from \textsf{PSRDISP} shown in black (top panel), and the residual DM time series after removing the determinsitic SW effects (bottom panel).}
    \label{fig:rev-DM-det-dmn-arn-swn-SWcomp-narrowband}
\end{figure*}

In this section, we discuss the details of the simulation-based injection and recovery study performed to validate the described technique. \update{As mentioned in Section \ref{sec:2}, since this method is agnostic to the DM estimation methods, we only present a representative study with the narrowband paradigm here, without any loss of generality. Additionally, we present a representative example using the wideband paradigm in \ref{AppA} for completeness.}

\begin{table}[!ht]
\centering
\begin{tabularx}{\columnwidth}{lcX}
\noalign{\smallskip}\hline
\noalign{\smallskip}
\textbf{Parameter} & \textbf{Prior distribution} & \textbf{Description} \\
\noalign{\smallskip}\hline
\noalign{\smallskip}
\multicolumn{2}{l}{\textit{White Noise}}\\
\noalign{\smallskip}
\noalign{\smallskip}
\texttt{DISPEFAC} & $\mathcal{U}(0.1, 7)$ & \textsf{PSRDISP} scaling factor for $\sigma_{\rm DM}$\\
\noalign{\smallskip}\hline
\noalign{\smallskip}
\multicolumn{2}{l}{\textit{Dispersion Measure Noise}}\\
\noalign{\smallskip}
\noalign{\smallskip}
$\gamma_{\rm DMN}$ & $\mathcal{U}(0, 7)$ & \texttt{DMN} spectral index\\
$\log_{10}A_{\rm DMN}$ &  $\mathcal{U}(-16, -10)$ & \texttt{DMN} log-amplitude\\
\noalign{\smallskip}\hline
\noalign{\smallskip}
\multicolumn{2}{l}{\textit{Deterministic Solar Wind}}\\
\noalign{\smallskip}
\noalign{\smallskip}
\texttt{NE\_SW} & $\mathcal{U}(0, 20)$ & Constant SW electron density\\
\texttt{NE\_SW1} & $\mathcal{U}(-10, 10)$ & First order time-derivative of \texttt{NE\_SW}\\
\texttt{NE\_SW2} & $\mathcal{U}(-10, 10)$ & Second order time-derivative of \texttt{NE\_SW} \\
\noalign{\smallskip}\hline
\noalign{\smallskip}
\multicolumn{2}{l}{\textit{Solar Wind Noise}}\\
\noalign{\smallskip}
\noalign{\smallskip}
$\gamma_{\rm SWN}$ & $\mathcal{U}(-4, 5)$ & \texttt{SWN} spectral index\\
$\log_{10}A_{\rm SWN}$ &  $\mathcal{U}(-9, 0)$ & \texttt{SWN} log-amplitude\\
\noalign{\smallskip}\hline
\noalign{\smallskip}
\end{tabularx}
\caption{The prior distributions used for the parameters of various deterministic and stochastic noise processes. The units of \texttt{NE\_SW}, \texttt{NE\_SW1} and \texttt{NE\_SW2} are $\rm cm^{-3}$, $\rm cm^{-3}\,\rm yr^{-1}$ and $\rm cm^{-3}\,\rm yr^{-2}$, respectively, while all other quantities are dimensionless. The different priors used for \texttt{SWN} hyper-parameters are based on the implementation in \textsf{ENTERPRISE}, and as discussed in \citet{Susarla+2024}. The symbol $\mathcal{U}(\alpha,\beta)$ stands for a Uniform distribution between $\alpha$ and $\beta$. Column 3 gives a description of the listed parameters for different processes.}
\label{tab:prior-distributions}
\end{table}

\cprotect\update{We used the \textsf{PINT} package to simulate narrowband ToAs with a span of $\sim$$20\,\rm yr$ at a cadence of 14 days, with $\sigma_{\rm ToA}=1\mu s$ and 16 sub-bands per epoch in the $300-500\,\rm{MHz}$ observing frequency range. We have taken the EPTA \texttt{DR2Full+} \citep{Antoniadis+2023} pulsar ephemeris for PSR J1909$-$3744 as the starting timing solution. We consider the \texttt{DMEPOCH} to be the middle of the data span, without any loss of generality. We included the deterministic DM and SW processes (up to the second order), with $\verb|NE_SW|=2.0$, $\verb|NE_SW1|=1.0$, and $\verb|NE_SW2|=0.75$ for the SW process, while the DM process is fixed as per the original timing solution. We injected \texttt{DMN} with $\log_{10}A=-13.5$ and $\gamma=3.0$. We included additional 16 frequency bins below the $1/T_{\rm span}$ bin, with a scale factor of 2, such that the lowest frequency bin corresponds to $1/2^{16}T_{\rm span}$ for the red noise process. We considered $N_{\rm harm}=1000$ for the injections, such that the high-frequency cutoff corresponds to $1000/T_{\rm span}$. This choice of Fourier basis ensures that the red noises are close to reality. Furthermore, we used $\verb|EFAC|=1.2$ for white noise\footnote{The parameters \texttt{EFAC} and \texttt{DMEFAC} are the conventional white-noise parameters \citep{LentatiAlexander+2014, Luo+2021, Susobhanan+2024, Susobhanan+vanHaasteren2025}, while \texttt{DISPEFAC} is the white-noise parameter introduced in the present work (see Section \ref{sec:2} for details).}. The simulated narrowband ToA residuals are shown in Figure \ref{fig:ref-DM-det-dmn-arn-swn-injections-narrowband}. To estimate the DM time series from these sub-banded ToAs, we used a modified version of the \textsf{DMcalc} package \citep{Krishnakumar+2021} which works on the ToAs directly instead of its original profile-domain implementation. The median precision of the DMs is estimated to be $\sim4.6\times10^{-5}\,\dmu$ and the extracted DMs are shown in Figure \ref{fig:rev-DM-det-dmn-arn-swn-realisations-narrowband} (as crimson points).}

\cprotect\update{We applied the \textsf{PSRDISP} technique on the extracted DMs and estimated the parameters of the dispersive noise processes, employing the priors listed in Table \ref{tab:prior-distributions}. We consider $N_{\rm harm}=100$ for recovery in the reduced-rank approximation \citep{vanHaasteren+Vallisneri2014} for the GP. In order to reconstruct the time-domain realisations of the process, we draw samples from a multivariate Gaussian distribution centered at the median values of the parameter posterior distributions and having a width, $\sigma_{\rm realisation}=3\sigma_{\rm posterior}$, for each parameter. We plot 20 such random realisations along with the extracted DM time series in Figure \ref{fig:rev-DM-det-dmn-arn-swn-realisations-narrowband}, and against the injected DM noise realisation (in the ToAs) in Figure \ref{fig:rev-DM-det-dmn-arn-swn-realisations-injection-narrowband}. It can be clearly seen that the reconstruction closely resembles the DM time series and the injected noise process, thereby proving the efficacy of the recovery via \textsf{PSRDISP}. The posterior distributions of the parameters are shown in Figure \ref{fig:rev-DM-det-dmn-arn-swn-noise-narrowband}, where the \texttt{DMN} amplitude is recovered within $1\sigma$ level, but the spectral index is recovered at $\sim$$2\sigma$ level. However, the $\sim$$3\sigma$ recovery of $\verb|NE_SW|$ and $\verb|NE_SW2|$ can be attributed to the strong covariance between them, and doesn't affect the generated realisations. This can also be assessed from the bottom panel in Figure \ref{fig:rev-DM-det-dmn-arn-swn-SWcomp-narrowband}, which is obtained after subtracting the determinisitic processes from the extracted DM time series, wherein no prominent residual SW effects are visible, apart from the stochasticity from \texttt{DMN}. \cprotect\newupdate{Since PSR J1909$-$3744 has an ecliptic latitude of about $-15^\circ$, we reckon that the small value of $\verb|NE_SW|$ couldn't be reliably distinguished from the second order term by its effect on the DMs, therefore giving large covariances in the estimated posteriors. This does not reflect any issue with our implemented technique in particular.} We can, therefore, infer that the recovery is consistent with the injections.}

\section{Discussion} \label{sec:4}

\update{The results presented in the previous section highlight the effectiveness of the \textsf{PSRDISP} technique in estimating noise processes directly from the high-precision epoch-wise DMs. This technique is, therefore, complementary to the conventional SPNA approach implemented in packages such as \textsf{ENTERPRISE}, where both chromatic and achromatic noise processes are modeled via delays introduced to ToAs. This treatment makes the conventional approach susceptible to spectral leakages that may arise due to complicated interplay of multiple noise processes, especially when they share a common Fourier basis. On the other hand, in the wideband SPNA implementation presented in \citet{Susobhanan+vanHaasteren2025} and \citet{Susobhanan+2026}, this effect is reduced, as noise processes are jointly modeled over simultaneous ToAs and DMs. The latter method, therefore, leverages the advantage of epoch-wise DMs, to model chromatic effects in a way less prone to spectral leakages. That said, this technique is relatively new and it's usage in the pulsar timing community is rather limited.}

\update{The \textsf{PSRDISP} technique is limited to modeling processes having imprints on the DMs. Therefore, we can use this alternate technique to effectively complement the conventional (narrowband) SPNA by, for instance,  employing the posterior distributions of the parameters estimated using the \textsf{PSRDISP} technique, to generate empirical distributions that can effectively tune the Markov chains employed to sample the corresponding parameter space with the narrowband SPNA technique. We are currently investigating this prospect with simulations. Having said so, care should be taken as to not lead to circular analysis on grounds discussed in \citet{vanHaasteren2024}, for instance, by using the results of this technique as prior distributions for other SPNA approaches on the same dataset.}

\update{A peculiar issue is also posed by the presence of scattering-induced effects on the estimated DMs. As detailed in Section \ref{sec:1}, the scattering-induced profile-shape distortions need to be adequately corrected before the DMs can be subjected to further noise analysis with the \textsf{PSRDISP} technique. This limitation also exists for the wideband paradigm currently \citep{Pennucci2019}. While the narrowband SPNA can indeed account for generic chromatic processes \citep{Nobleson+2026, Larsen+2026}, the adequate propagation of their effects to the estimated DMs is still not well understood. Furthermore, it should be noted that the \textsf{PSRDISP} technique cannot be directly used to estimate achromatic noise processes, or perform PTA gravitational wave analyses. Rather, this technique provides a powerful validation tool to check the consistency of chromatic noise estimates derived from the conventional SPNA methods, which is of paramount importance to accurately characterise the single-pulsar noise budget of a PTA.}

\section{Summary} \label{sec:5}
In this work, we presented a novel approach, \textsf{PSRDISP}, to model dispersive single-pulsar noise processes in pulsar timing datasets, complementing the existing SPNA techniques. The conventional narrowband SPNA methodology, as implemented in the \textsf{ENTERPRISE} package, \update{models various noise processes as temporal perturbations to the original timing solution, assuming them to be significantly smaller than the period of the pulsar. The stochastic processes are modeled as Fourier-basis GPs, commonly with a power-law kernel. This framework, however, isn't resistant to the drawbacks of spectral leakages due to the interplay of various noise processes on a shared Fourier basis. As discussed in \citet{Iraci+2024} and \citet{Larsen+2024}, such contaminations could lead to sub-optimal DM noise extraction using GPs, which can significantly affect the sensitivity of a low observing-frequency PTA experiment to detect and accurately characterise the GWB}.

We addressed these potential issues by modeling the dispersive noise processes directly using high-precision epoch-wise DM estimates. The uniqueness of this approach is that these DMs are almost unaffected by leakages from the achromatic red noise, and to an extent, white noise processes. We implemented this technique using the Fourier-basis GP approach of \citet{LentatiAlexander+2014} for stochastic processes in the reduced-rank approximation. \update{This technique is inherently agnostic to the narrowband and wideband paradigms, as long as the epoch-wise DMs are mutually independent.}

In order to assess the efficacy of noise modeling using the developed technique, we presented a representative example wherein we simulated wideband DMs using the \textsf{PINT} package with realistic injections of noise processes. We estimated the parameter posteriors using the \textsf{emcee} package, which employs an affine-invariant ensemble sampling algorithm \citep{Foreman-Mackey+2013, GoodmanWeare2010}. Additionally, we reconstructed the time-domain realisations of the stochastic processes via a multivariate Gaussian distribution centered at the median parameter estimates, and having a width equivalent to $3\sigma$ level of the parameter posterior distributions. The recovered GP hyper-parameters were found to be $\lesssim$$1\sigma$ level of the injected values. The reconstructed time-domain realisations closely resembled the simulated DM time series. \update{We have also presented an additional case study using \newupdate{wideband} simulations in \ref{AppA}, where again, we get good recovery using the developed technique.} 

\update{These simulations demonstrated the efficacy of noise extraction using high-precision epoch-wise DMs, which can serve as a complementary technique to the conventional ToA-based noise analysis methodologies. Therefore, this work holds immense potential in validating the existing SPNA techniques by providing an alternative framework to model dispersive single-pulsar noises.}

\section{Future Work} \label{sec:6}
As outlined in Section \ref{sec:4}, we are currently investigating the prospect of complementing the conventional SPNA techniques with the framework discussed in the present work. We are also planning to incorporate spatially correlated dispersive processes, such as a common solar wind process, to account for correlated DM effects in the ensemble of pulsars observed by a PTA. The prospect of addressing the effect of scattering-induced variations over DMs and their modeling in this framework is also being explored.

\section*{Acknowledgements}
The work of CD at the Physical Research Laboratory (PRL) was supported by the Department of Space, Government of India. CD acknowledges the Param Vikram-1000 High Performance Computing Cluster of the Physical Research Laboratory (PRL), a unit of the Department of Space, Government of India, for performing the intensive computations. The authors acknowledge Bhal Chandra Joshi for his insights and valuable comments. CD acknowledges RM for the immense support during the work.

\section*{Software}
\begin{itemize*}
    \item \textsf{PYTHON} \citep{vanRossum+2009}
    \item \textsf{PINT} \citep{Luo+2021, Susobhanan+2024}
    \item \textsf{DMcalc} \citep{Krishnakumar+2021}
    \item \textsf{TEMPO2} \citep{Edwards+2006}
    \item \textsf{emcee} \citep{Foreman-Mackey+2013}
    \item \textsf{ASTROPY} \citep{Whelan+2022}
    \item \textsf{SCIPY} \citep{Virtanen+2020}
    \item \textsf{MATPLOTLIB} \citep{Hunter2007}
    \item \textsf{CORNER} \citep{Foreman-Mackey2016}
    \item \textsf{NUMPY} \citep{Harris+2020}
\end{itemize*}

\section*{Data Availability}
The entire work presented in this paper is based on simulated datasets, generated solely for this purpose using publicly available software. The datasets are also shared along with the manuscript as supplementary material.

\bibliographystyle{elsarticle-harv} 
\bibliography{PSRDISP_revision}

\clearpage
\appendix
\onecolumn

\section{A representative example using the wideband paradigm} \label{AppA}
\setcounter{figure}{0}

\begin{figure*}[!ht]
\centering
    \subfigure[DM time series]{\includegraphics[trim=0 0 0 0, clip, width=0.49\textwidth]{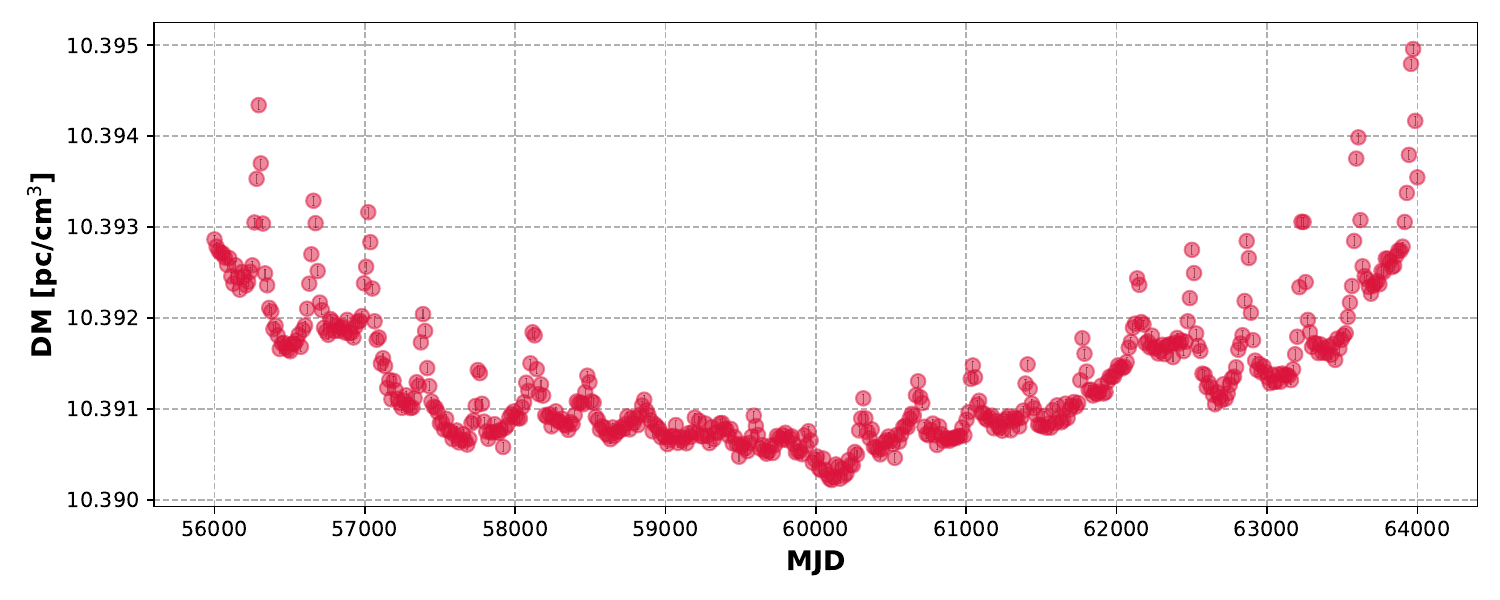}}
    \subfigure[ToA residuals]{\includegraphics[trim=0 0 0 0, clip, width=0.49\textwidth]{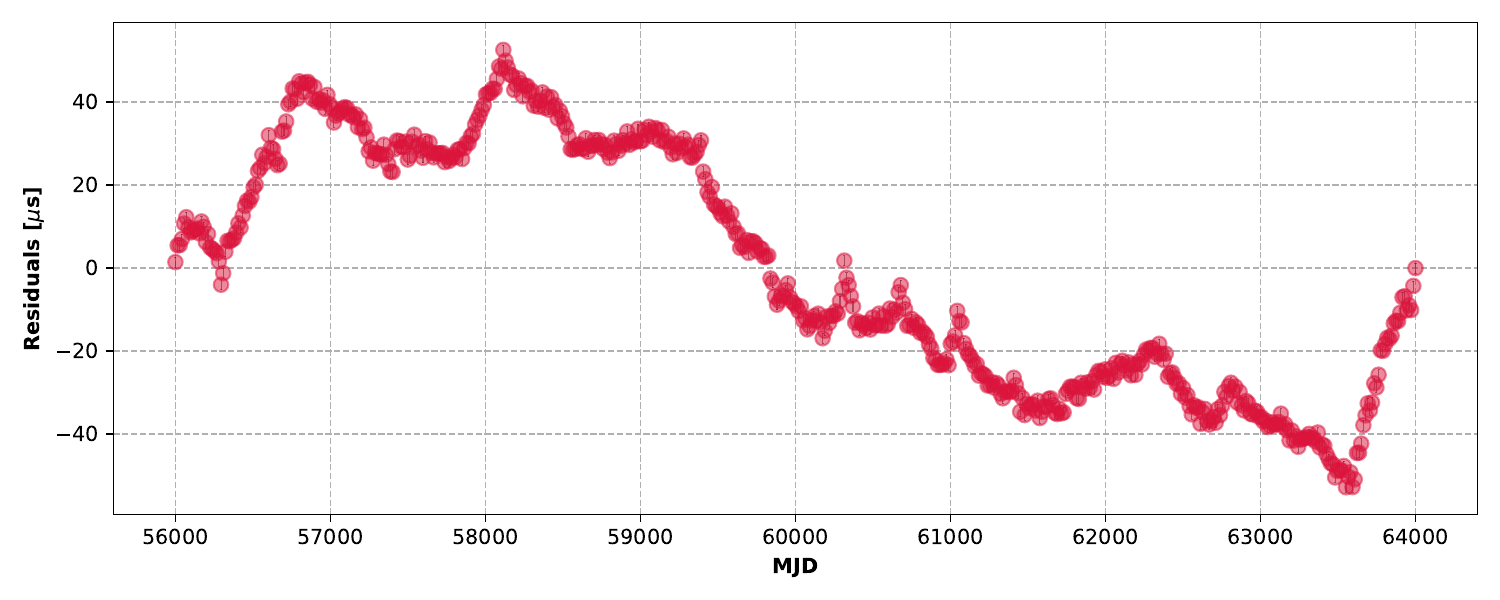}}
\cprotect\caption{The simulated wideband DMs (left) and ToA residuals (right). The DMs have a sensitivity of $5\times10^{-5}\,\dmu$, while that of ToAs is $1\mu\rm s$. The deterministic SW process is simulated with $\verb|NE_SW|=2.0$, $\verb|NE_SW1|=0$, and $\verb|NE_SW2|=1.0$, while the DM process is set as per the original timing solution. The \texttt{DMN} process is injected with $\log_{10}A=-13.5$ and $\gamma=3.0$, \texttt{ARN} with $\log_{10}A=-13.0$ and $\gamma=3.5$ and \texttt{SWN} with $\log_{10}A=-6.8$ and $\gamma=2.5$. Additional 16 Fourier bins below the $1/T_{\rm span}$ bin are incorporated to simulate a realistic noise process. The high-frequency cutoff is set to $1000/T_{\rm span}$.}
\label{fig:AppA1}
\end{figure*}

\cprotect\update{For this representative case, we simulated wideband ToAs and DMs with the deterministic DM and SW processes (up to the second order), along with the achromatic red noise\footnote{This process, as suggestive by its name, does not have any chromatic effects. Such a process is modeled by a power-law PSD given by equation \ref{eq:2.15} with $\chi=0$.} (\texttt{ARN}), \texttt{DMN} and \texttt{SWN} processes. We injected $\verb|NE_SW|=2.0$, $\verb|NE_SW1|=0$, and $\verb|NE_SW2|=1.0$ for the deterministic SW process, while the DM process is fixed as per the original timing solution. We injected \texttt{DMN} with $\log_{10}A=-13.5$ and $\gamma=3.0$, \texttt{ARN} with $\log_{10}A=-13.0$ and $\gamma=3.5$ and \texttt{SWN} with $\log_{10}A=-6.8$ and $\gamma=2.5$. The red noise basis is defined as per Section \ref{sec:3}. Furthermore, we used $\verb|DMEFAC|=\verb|EFAC|=1.2$ for white noise. The simulated wideband DMs and ToA residuals are shown in Figure \ref{fig:AppA1}.}

\begin{figure*}[!ht]
    \centering
    \includegraphics[trim=0 0 0 0, clip, width=\textwidth]{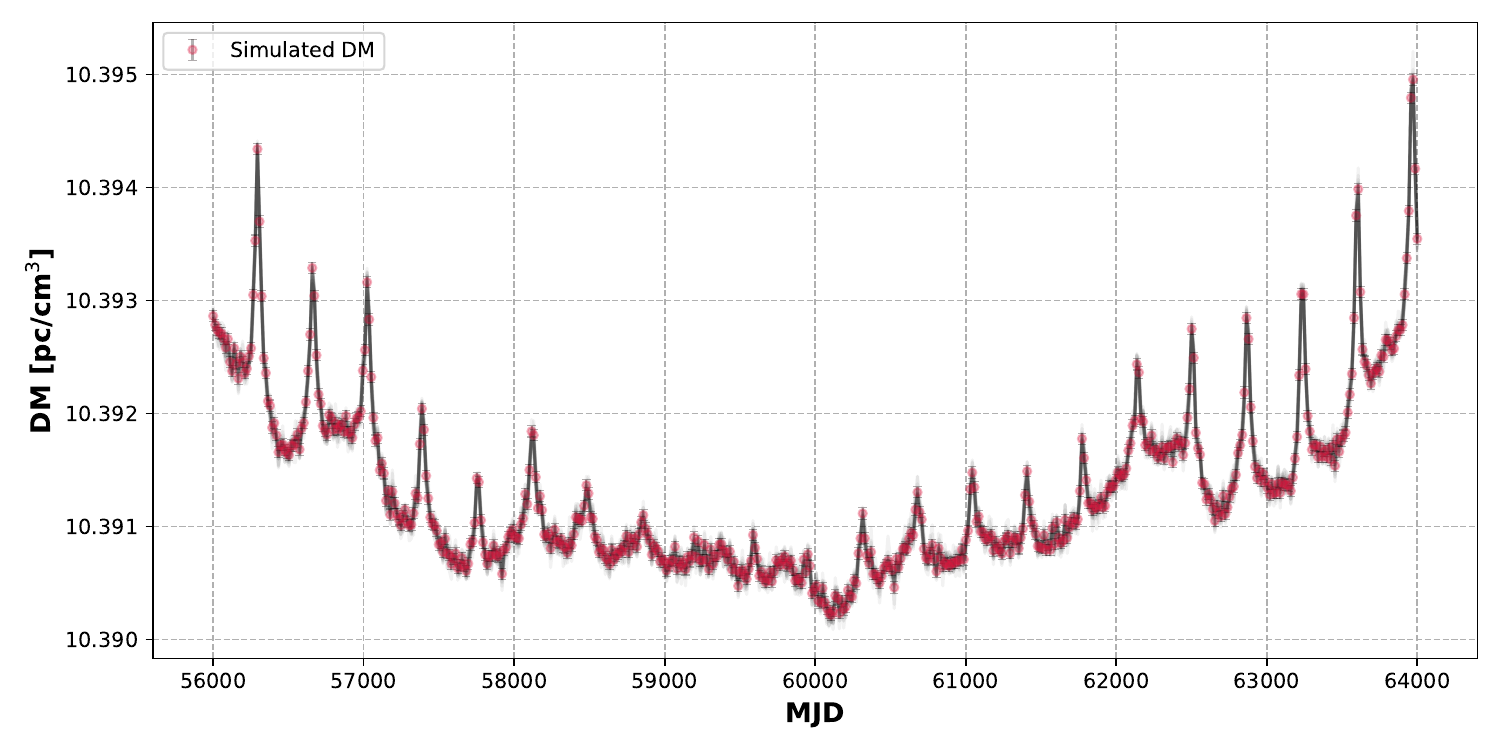}
    \caption{The simulated wideband DM time series along with 20 random time-domain reconstructed noise process realisations using parameter estimates obtained after applying the \textsf{PSRDISP} technique on the DM time series (shown in solid black).}
    \label{fig:AppA2}
\end{figure*}

\begin{figure*}[!ht]
    \centering
    \includegraphics[trim=0 0 0 0, clip, width=\textwidth]{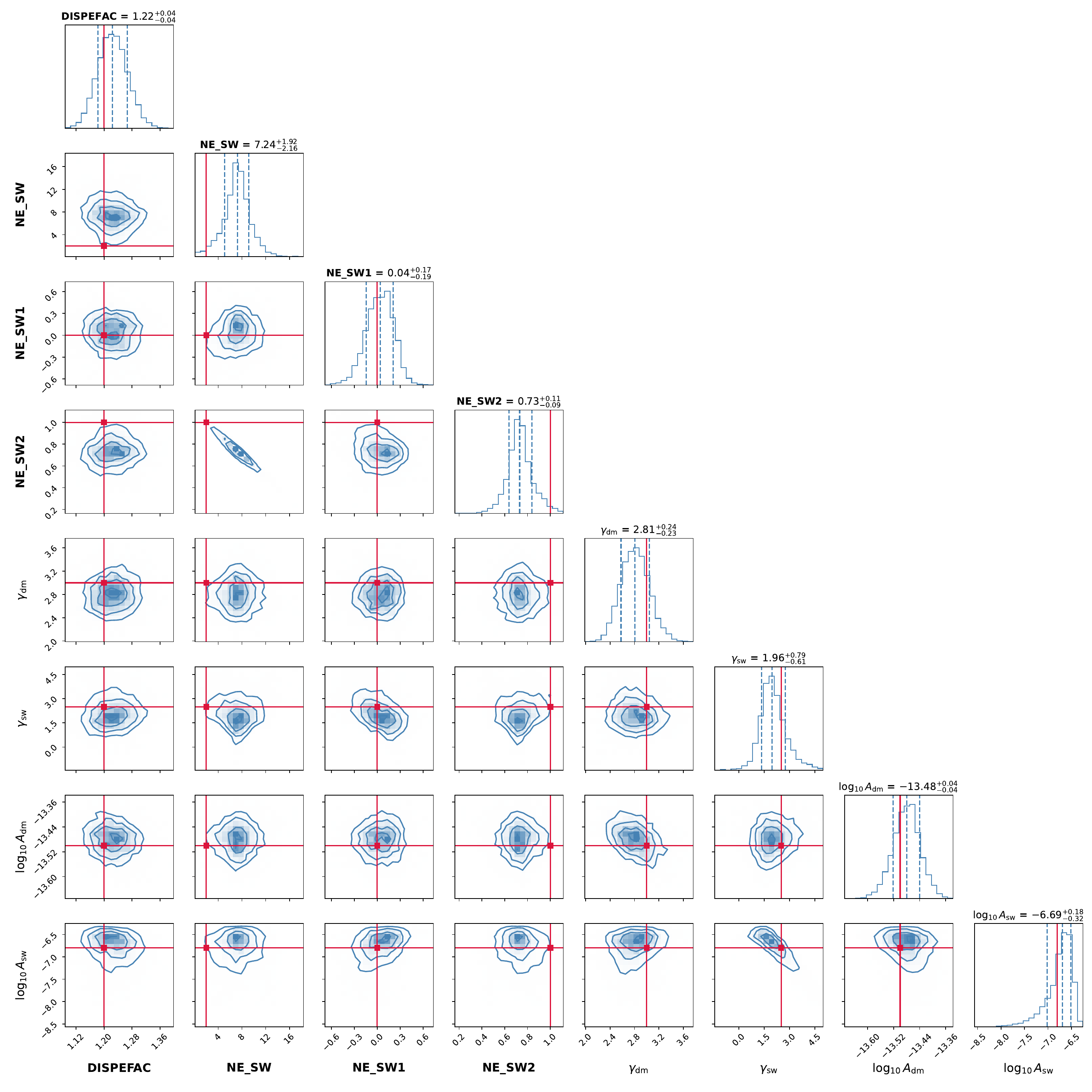}
    \cprotect\caption{The posterior distributions of deterministic SW process and the DM \texttt{WN} parameters, along with \texttt{DMN} and \texttt{SWN} hyper-parameters, estimated using the \textsf{PSRDISP} technique on the simulated wideband DM time series. The injected values of various parameters are highlighted as crimson vertical lines, while the median values along with $1\sigma$ ranges are shown at the top of each posterior panel.}
    \label{fig:AppA3}
\end{figure*}

\begin{figure*}[!ht]
    \centering
    \includegraphics[trim=0 0 0 0, clip, width=\textwidth]{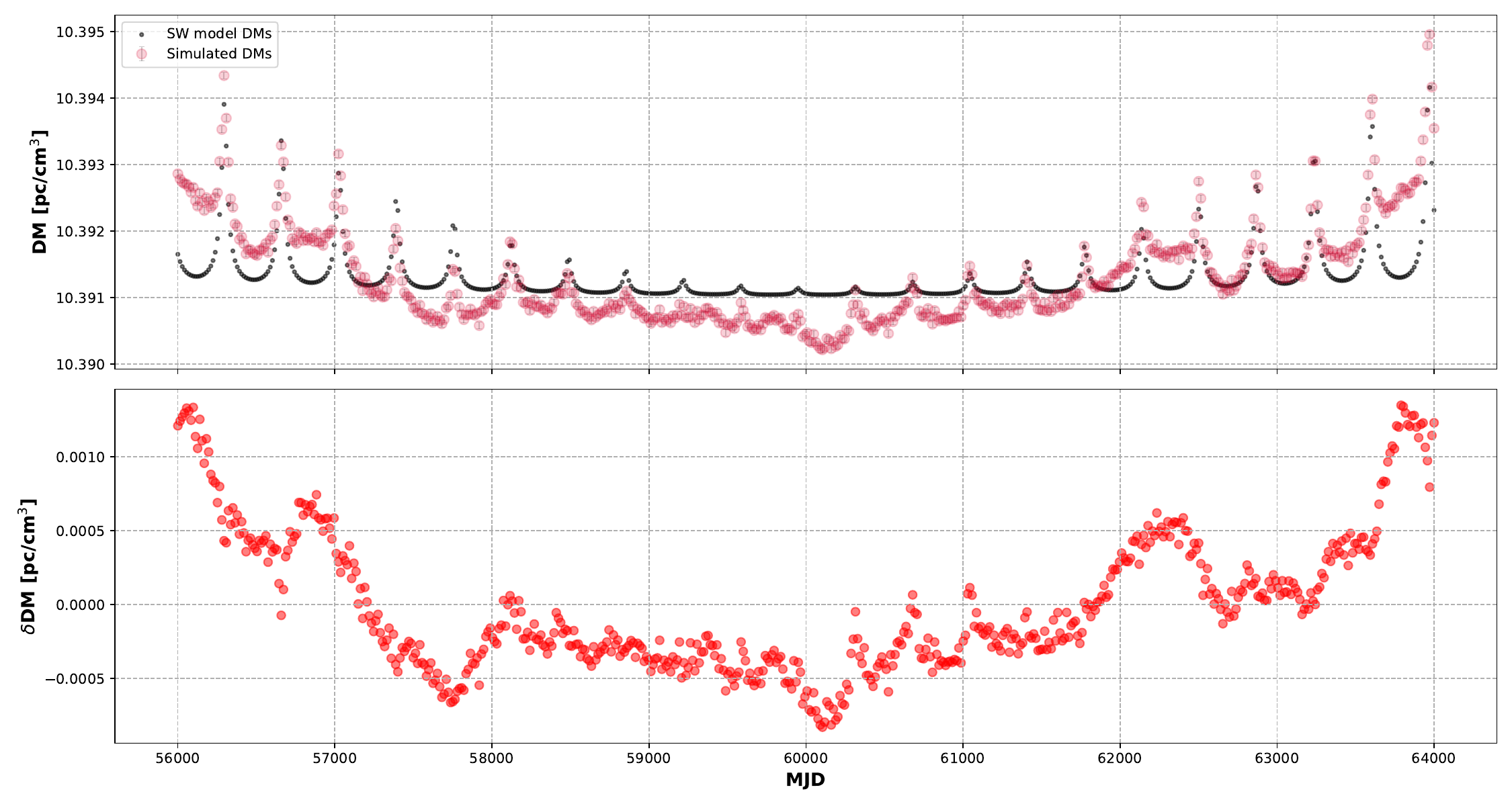}
    \cprotect\caption{The simulated wideband DM time series along with the deterministic SW process estimated from \textsf{PSRDISP} shown in black (top panel), and the residual DM time series after removing the determinsitic SW effects (bottom panel).}
    \label{fig:AppA4}
\end{figure*}

\cprotect\update{We applied the \textsf{PSRDISP} technique, on the simulated DMs, and estimated the parameters of the dispersive noise processes, employing the priors listed in Table \ref{tab:prior-distributions} with $N_{\rm harm}=100$ for all the dispersive GPs. The time-domain realisations are constructed as described in Section \ref{sec:3}. We plot 20 such random realisations along with the simulated DM time series in Figure \ref{fig:AppA2}. It can be clearly seen that the reconstruction closely resembles the actual simulated DMs, thereby proving the efficacy of the recovery via the \textsf{PSRDISP} technique.}

\cprotect\update{The posterior distributions of the parameters are shown in Figure \ref{fig:AppA3} along with the injected values as crimson vertical lines in the respective posterior panels. It is evident that all the GP hyper-parameters are recovered within $1\sigma$ level of their injected values. However, the $\sim$$3\sigma$ recovery of $\verb|NE_SW|$ and $\verb|NE_SW2|$ can be attributed to the strong covariances, and can assessed from the bottom panel in Figure \ref{fig:AppA4}, which is obtained after subtracting the determinisitic DM and SW process from the simulated DM time series, wherein no prominent residual SW effects are visible, apart from the residual stochasticity due to \texttt{DMN} and \texttt{SWN}. The recovery is, therefore, consistent with the injections.}

\update{It is worthwhile, however, to point out that the \texttt{DISPEFAC} value is close to the injected \texttt{DMEFAC} value in this case, which is not necessarily the case with narrowband and can be seen from Figure \ref{fig:rev-DM-det-dmn-arn-swn-noise-narrowband}. This is in line with the arguments presented in the beginning of Section \ref{sec:3} as well.}

\end{document}